\documentclass[11pt,onecolumn,aps,superscriptaddress,
               amsmath,amssymb,nofootinbib]{revtex4}      
\usepackage{graphicx}

\usepackage{graphicx,epsfig}

\usepackage{dcolumn}
\usepackage{bm}

\newcommand{\etal} {\mbox{$et~al.$}}

\begin{document}

\begin{flushright}
\baselineskip=12pt \normalsize
{ACT-05-08},
{MIFP-08-23}\\
\smallskip
\end{flushright}

\title{Supergravity and Superstring Signatures of the One-Parameter Model at LHC}

\author{James A. Maxin}
\affiliation{George P. and Cynthia W. Mitchell Institute for
Fundamental Physics, Texas A\&M
University,\\ College Station, TX 77843, USA}
\author{Van E. Mayes}
\affiliation{George P. and Cynthia W. Mitchell Institute for
Fundamental Physics, Texas A\&M
University,\\ College Station, TX 77843, USA}
\author{Dimitri V. Nanopoulos}
\affiliation{George P. and Cynthia W. Mitchell Institute for
Fundamental Physics, Texas A\&M
University,\\ College Station, TX 77843, USA}
\affiliation{Astroparticle Physics Group, Houston
Advanced Research Center (HARC),
Mitchell Campus,
Woodlands, TX~77381, USA; \\
Academy of Athens,
Division of Natural Sciences, 28~Panepistimiou Avenue, Athens 10679,
Greece}

\begin{abstract}
\begin{center}
{\bf ABSTRACT}
\end{center} 
Many string constructions have a classical no-scale structure, resulting in a one-parameter model (OPM) for the supersymmetry breaking soft terms. As a highly constrained subset of mSUGRA, the OPM has the potential to be predictive. Conversely, if the observed superpartner spectrum at LHC is a subset of the OPM parameter space, then this may provide a clue to the underlying theory at high energies.  We investigate the allowed supersymmetry parameter space for a generic one-parameter  model taking into account the most recent experimental constraints.  We find that in the strict moduli scenario, there are no regions of the parameter space which may satisfy all constraints. However, for the
dilaton scenario, there are small regions of the parameter space where all constraints may
be satisfied and for which the observed dark matter density may be generated.  We also survey the possible signatures which may be observable at the Large Hadron Collider (LHC).  Finally, we compare collider signatures of OPM to those from a model with non-universal soft terms, in particular those of an intersecting $D6$-brane model. We find that it may be possible to distinguish between these diverse scenarios at LHC. 
\end{abstract}

\maketitle

\newpage
\section{Introduction}

With the dawn of the 
Large Hadron Collider (LHC)
era, the prospects for the discovery of new physics 
may finally be arriving. 
In particular, whatever physics is responsible for stabilizing 
the electroweak scale should be discovered.  Signals of
the favored mechanism, broken supersymmetry, may be observed as 
well as the Higgs states required to break the electroweak symmetry. 
However, at present there is no theory which may uniquely predict the masses
of the superpartners should they be observed at LHC

In principle, it should be possible to 
derive all known physics in a top-down 
approach directly from 
a more fundamental theory such as string theory, 
as well as potentially 
predicting new and unexpected phenomena. 
Conversely, following a bottom-up approach, one may ask
if it is possible to deduce the origin of new
physics given such a signal at LHC.  
For example, in the case of low-energy supersymmetry,
it may be possible from the experimental data to deduce 
the structure of the fundamental theory at high energy scales
which determines the
soft-supersymmetry breaking terms and ultimately leads
to radiative electroweak symmetry breaking (REWSB)~\cite{Ellis:1982wr, Ellis:1983bp, AlvarezGaume:1983gj}.  

No-scale supergravity (nSUGRA)~\cite{DVNS} is such a framework where it is possible to naturally 
explain REWSB and correlate it with the gravitino mass, or more generally, the effective SUSY breaking scale. In the simplest no-scale models, the gravitino mass
$m_{3/2}$ remains undetermined at the classical level, and is instead fixed
by radiative corrections to be near the electroweak scale~\cite{DVNS}. Thus, in this framework,
we find that the scale of supersymmetry breaking is correlated with the electroweak scale~\cite{DVNS}. Another striking feature of nSUGRA is that the cosmological constant vanishes at tree-level.  
Although it is presently known that the cosmological constant is in fact non-zero, its very small value is still consistent with the no-scale framework with small corrections.
Furthermore, it is well-known that the K\a"ahler moduli of Type I, IIB
orientifold, 
and heterotic string compactifications have a classical no-scale structure~\cite{JLDN, VKJL, ABIM}.  It has been shown that in Type IIB orientifold compactifications, 
this type of supersymmetry breaking corresponds to turning on RR and NS fluxes, which 
are generically present in order to cancel tadpoles as well as to stabilize closed-string
moduli.  Indeed, this is the case in the so-called large volume models~\cite{BBCQ}~\cite{JCQS}. This combined with the generic appearance of the 
no-scale structure across many string compactifications leads to the idea that supersymmetry
breaking is moduli dominated.

In string models, supersymmetry breaking is typically performed in a hidden sector as well as 
through the universal moduli and dilaton fields.  For a given string compactification, the precise nature of the
supersymmetry breaking is determined by model-dependent calculations.  However, at
present there are no specific string compactifications which completely satisfy all theoretical
criteria which are desired in such a model.  Thus, a model-independent approach is perhaps wiser at the present time.  The most studied model of supersymmetry breaking is minimal supergravity (mSUGRA), which arises
from adopting the simplest ansatz for the K\a"ahler metric, treating all chiral superfields
symmetrically.  In this framework,~~$\mathcal{N}=1$ supergravity is broken in a hidden sector
which is communicated to the observable sector through gravitational interactions.  Such models
are characterized by the following parameters: a universal scalar mass $m_0$, a universal gaugino mass
$m_{1/2}$, the Higgsino mixing $\mu$-parameter, the Higgs bilinear $B$-parameter, a universal trilinear coupling $A_0$, and tan~$\beta$.  One then determines the $B$ and $|\mu|$ parameters by the minimization of the Higgs potential triggering REWSB, with the sign of $\mu$ remaining undetermined.  Thus, we are left with only four parameters.  Although,
this is one of the most generic frameworks that can be adopted, and many string compactifications
typically yield expressions for the soft terms which are even more constrained due to the 
no-scale structure which emerges naturally in these theories assuming that supersymmetry breaking is dominated by the K\a"ahler moduli and/or dilaton.  In particular, in such nSUGRA 
models, we will generically have $m_{0} = m_{0}(m_{1/2})$ and $A = A(m_{1/2})$. This reduces the number of free parameters compared to mSUGRA down to two, $m_{1/2}$ and tan$\beta$.  In fact, adopting a 
strict no-scale framework, one can also fix the $B$-parameter as $B=B(m_{1/2})$, and thus we
are led to a {\it one-parameter} model where all of the soft terms may be fixed in terms of 
$m_{1/2}$.

If we assume that the supersymmetry breaking is triggered by some of the moduli fields in a given string compactification, namely the dilaton $S$ and the three K\a"ahler moduli $T$ which obtain VEVs 
$\left\langle F_S \right\rangle$ and $\left\langle F_T \right\rangle$ respectively, a simple expression for the scalar masses may be adopted:
\begin{equation}
\tilde{m}^2_i = m^2_{3/2} (1 + n_i \mbox{cos}^2 \theta), 
\end{equation}
with $\mbox{tan}\theta = \left\langle F_S \right\rangle / \left\langle F_T\right\rangle$ and where $m_{3/2}$ is the gravitino mass and $n_i$ are the modular weights of the respective 
matter fields.

In order to obtain universal scalar masses, which are highly suggested by the required absence of FCNC~\cite{Ellis:1981ts}, there are two possible cases which may 
be considered:  (i)  setting $\theta = \pi/2$ so that 
$\left\langle F_S \right\rangle >> \left\langle F_T\right\rangle$; or (ii) setting
all $n_i$ to be the same ($n_i=-1$) and $\theta=0$ so that $\left\langle F_T \right\rangle >> \left\langle F_S\right\rangle$ so that all scalar masses vanish at the unification scale.  The first
of these two cases is referred to as the special {\it dilaton} scenario,
\begin{equation}
\label{eqn:dilaton}
m_{0} = \frac{1}{\sqrt{3}}m_{1/2}, \ \ \ \ \ A = -m_{1/2}, \ \ \ \ \ B = \frac{2}{\sqrt{3}}m_{1/2}.
\end{equation}
while the second is referred to as the strict {\it moduli} scenario,
\begin{equation}
\label{eqn:moduli}
m_{0} = 0, \ \ \ \ \ A = 0, \ \ \ \ \ B = 0.
\end{equation}

For many string compactifications, especially those within the free-fermionic class of models in particular those with a flipped $SU(5)$ gauge group~\cite{AEHN}, the soft-terms will have such a form.  Interestingly, soft terms for heterotic M-theory compactifications with moduli dominated supersymmetry breaking take
the form~\cite{TJLI}
\begin{eqnarray}
&m_{1/2} = \frac{x}{1+x}m_{3/2} \\ \nonumber
&m_{0} = \frac{x}{3+x}m_{3/2} \\ \nonumber
&A = - \frac{3x}{3+x}m_{3/2} 
\end{eqnarray}

\noindent while for dilaton dominated supersymmetry breaking they take the form
\begin{eqnarray}
&m_{1/2} = \frac{\sqrt{3}m_{3/2}}{1+x} \\ \nonumber
&m_{0}^{2} = m_{3/2}^{2} - \frac{3m_{3/2}^{2}}{(3+x)^{2}}x(6+x) \\ \nonumber
&A = - \frac{\sqrt{3}m_{3/2}}{3+x}(3-2x) 
\end{eqnarray}

\noindent which reduce to the above moduli and dilaton scenarios in the limit $x \rightarrow 0$,
where
\begin{equation}
x \propto \frac{(T+\overline{T})}{S+\overline{S}}
\end{equation}

In addition, the so-called large-volume models have been studied extensively~\cite{BBCQ}~\cite{JCQS} in recent years and the generic soft terms for this framework have been calculated in~\cite{CAQS}. These models involve Type IIB compactifications where the moduli are stabilized by fluxes and quantum corrections to the K\a"ahler potential generate an exponentially large volume. This exponentially large volume may lower the string scale to an intermediate level which can be in the range $m_{s} \sim 10^{3-15}$ GeV. In such models, the
soft terms can take the form
\begin{eqnarray}
 \centering
	&m_0 = \frac{1}{\sqrt{3}}M \nonumber \\
	&A_0 = -M \nonumber \\
	&B = - \frac{4}{3}M
\end{eqnarray}
where $M$ is a universal gaugino mass,
which are essentially identical to the special dilaton scenario given above. However, this framework
is different from our analysis in that the string scale may be lower than
what is usually taken for the grand unification scale $\approx 2.1 \times 10^{16}$~GeV.  Thus, in
this scenario the observed unification of the gauge couplings when extrapolated to
high energies is merely coincidental.  Of course, this then also affects the running
of the soft-masses resulting in different superpartner spectra than what would otherwise
be obtained.  More generally, no-scale moduli-dominant scenarios of SUSY breaking are
favored by F-theory~\cite{Aparicio:2008wh}, as are models with a flipped SU(5) gauge group~\cite{Beasley:2008kw}.  

In this work, we identify the regions of the supersymmetry parameter space for a generic one-parameter model which are allowed by current experimental constraints and survey the signatures which may be observable at LHC. 
We find that in the strict moduli scenario,
there are no regions of the parameter space which may satisfy all constraints. However, for the
dilaton scenario, there are small regions of the parameter space where all constraints may
be satisfied and for which the observed dark matter density may be generated.  The model is
thus a highly constrained subset of mSUGRA, which allows the model to potentially be predictive. Conversely, if
the superpartner spectrum actually observed at LHC lies within the OPM parameter space, then this may provide a strong clue
to the underlying theory at high energy scales.   Finally, we simulate
the different LHC signatures for this model and compare
to those for an intersecting $D6$-brane model which possesses many desirable phenomenological characteristics~\cite{RIBM}. We find certain signatures may indicate there are distinguishing phenomenological characteristics between these different types of constructions.

\section{Allowed Parameter Space of OPM}
  
A one-parameter model of the above form has been much studied in the past~\cite{Lopez:1993rm,Lopez:1994fz,Lopez:1995hg}.  However, the last such analysis was
performed some years ago.  In the intervening time,  the experimental constraints
on SUSY models have been updated considerably, especially in regards to the constraints
on the dark matter density.  In addition, the experimental determination of the top
quark mass has become considerably more precise in recent years.  Here, we will generate a set of soft terms at the unification scale using the ansatz given in Eqs.~\ref{eqn:dilaton} and
Eqs.~\ref{eqn:moduli} for both
the dilaton and moduli scenarios.  The soft terms are then input into {\tt MicrOMEGAs 2.0.7}~\cite{MCRO} using
{\tt SuSpect 2.34}~\cite{SUSP} as a front end to evolve the soft terms down to the electroweak scale via the Renormalization Group Equations (RGEs) and then to 
calculate the corresponding relic neutralino density.  We take the top quark mass to be
$m_t = 171.4$~GeV~\cite{TOPQ} and leave tan~$\beta$ as a free parameter, while $\mu$ is determined by
the requirement of REWSB. However, we do take $\mu > 0$ as suggested by the results of 
$g_{\mu}-2$ for the muon. The resulting superpartner spectra are filtered according to the following 
criteria:

\begin{enumerate}

\item The WMAP 5-year data~\cite{WMAP} for the cold dark matter density,  0.1109 $\leq \Omega_{\chi^o} h^{2} \leq$ 0.1177.  We also consider the WMAP 2$\sigma$ results~\cite{WMP3}, 0.095 $\leq \Omega_{\chi^o} h^{2} \leq$ 0.129. In addition, we look at the Supercritical String Cosmology (SSC) model~\cite{DNDX} for the dark matter density, in which a dilution factor of $\cal{O}$(10) is allowed~\cite{SSC2}, where $\Omega_{\chi^o} h^{2} \leq$ 1.1. For a discussion of the SSC model within the context of mSUGRA, see~\cite{Dutta:2008ge}. We investigate two cases, one where a neutralino LSP is the dominant component of the dark matter and another where it makes up a subdominant component such that
0 $\leq \Omega_{\chi^o} h^{2} \leq$ 0.1177, 0 $\leq \Omega_{\chi^o} h^{2} \leq$ 0.129, and 0 $\leq \Omega_{\chi^o} h^{2} \leq$ 1.1.

\item The experimental limits on the Flavor Changing Neutral Current (FCNC) process, $b \rightarrow s\gamma$. The results from the Heavy Flavor Averaging Group (HFAG)~\cite{HFAG}, in addition to the BABAR, Belle, and CLEO results, are: $Br(b \rightarrow s\gamma) = (355 \pm 24^{+9}_{-10} \pm 3) \times 10^{-6}$. There is also a more recent estimate~\cite{MMEA} of $Br(b \rightarrow s\gamma) = (3.15 \pm 0.23) \times 10^{-4}$. For our analysis, we use the limits $2.86 \times 10^{-4} \leq Br(b \rightarrow s\gamma) \leq 4.18 \times 10^{-4}$, where experimental and
theoretical errors are added in quadrature.

\item The anomalous magnetic moment of the muon, $g_{\mu} - 2$. For this analysis we use the lower bound $a_{\mu}$ $>$~ 11 $\times 10^{-10}$~\cite{MUON}.

\item The process $B_{s}^{0} \rightarrow \mu^+ \mu^-$ where the decay has a $\mbox{tan}^6\beta$ dependence. We take the upper bound to be $Br(B_{s}^{0} \rightarrow \mu^{+}\mu^{-}) < 5.8 \times 10^{-8}$~\cite{AAEA}.

\item The LEP limit on the lightest CP-even Higgs boson mass, $m_{h} \geq 114$ GeV~\cite{HIGG}.

\end{enumerate}

A scan of the full parameter space is performed for both the strict moduli scenario and the dilaton scenario. The gaugino mass $m_{1/2}$ is varied in increments of $1$~GeV in the range $50-2000$~GeV while tan$\beta$ is varied in increments of $0.1$ in the range $1-60$. 
For the moduli scenario for $m_{1/2},~~ m_{0} = 0,~~ A_{0} = 0,$ and tan$\beta$ taken as a free parameter, it is found that there are no spectra which satisfy all constraints.   This analysis was conducted for the strict no-scale moduli scenario only. However, solutions may potentially be found when non-leading order corrections to the no-scale model are taken into account. For a detailed study concerning these corrections to the no-scale model, see~\cite{Dutta:2007xr}. We conclude that there are no solutions for the moduli scenario unless these corrections are incorporated, and for the present work no further study will be conducted into the strict moduli scenario. 

\begin{figure}
  \centering
		\includegraphics[width=1.0\textwidth]{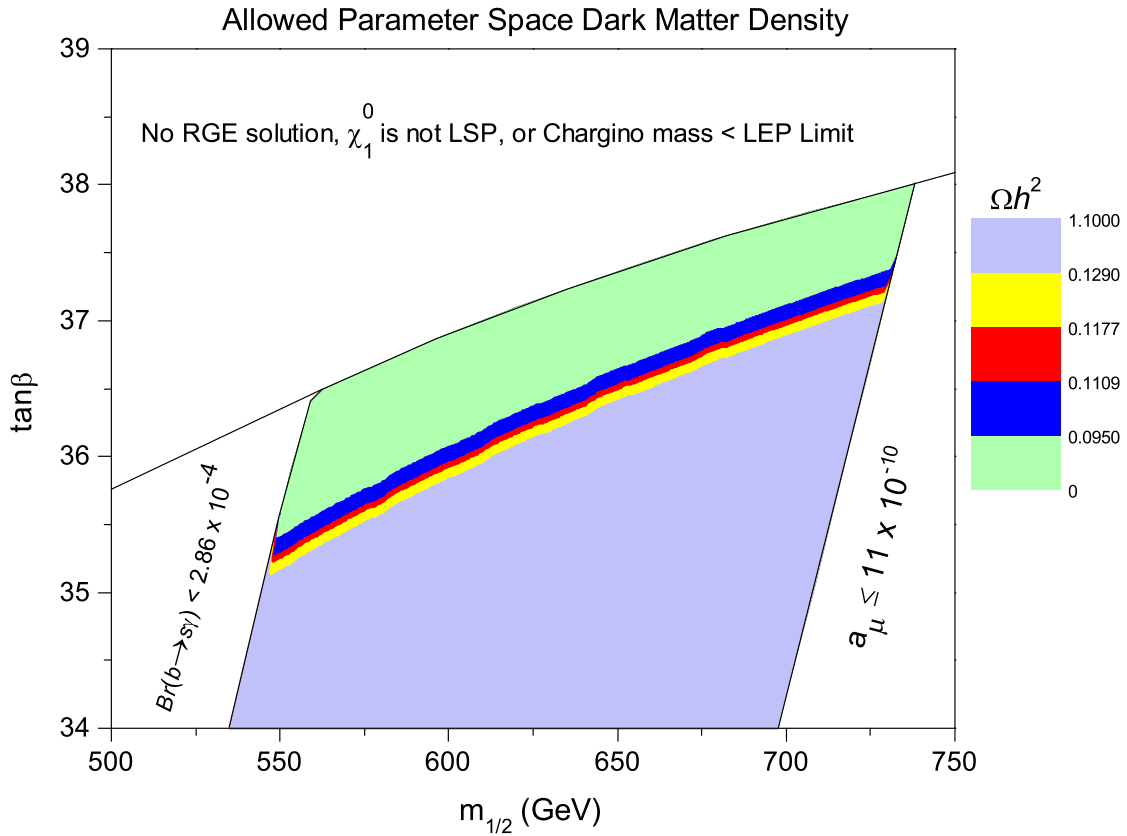}
		\caption{Parameter space allowed by all constraints for the WMAP constraints. The thin shaded areas constitute the WMAP region. The shaded area below the WMAP region is allowed by all constraints, though 0.129 $<$ $\Omega_{\chi^o} h^{2} \leq$ 1.1. The region on the far right side of the plot is excluded by the $g_{\mu}-2$ results. The region on the far left side of the plot is excluded by $Br(b \rightarrow s\gamma)$ $<$ $2.86 \times 10^{-4}$. The remaining area at the top of the plot is excluded for the reasons noted.}
	\label{fig:OPM_mgvstanb}
\end{figure}

\begin{figure}
  \centering
		\includegraphics[width=1.0\textwidth]{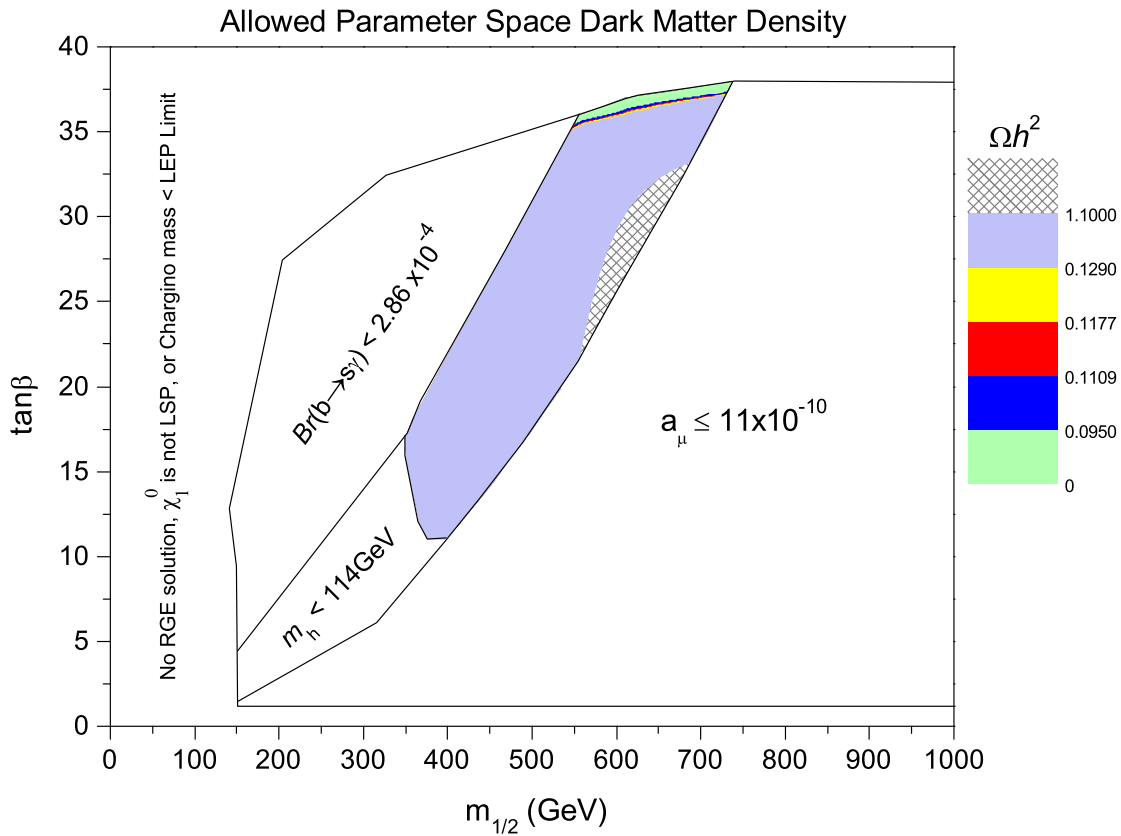}
		\caption{Parameter space allowed by all constraints for the SSC constraints. The WMAP region is the small thin region at the top. The shaded area below the WMAP region is allowed by all constraints, though 0.129 $<$ $\Omega_{\chi^o} h^{2} \leq$ 1.1. The region excluded by $m_{h}$ $<$ 114 GeV satisfies all other constraints, including $\Omega_{\chi^o} h^{2} \leq$ 1.1. The cross-hatched region satisfies all constraints, though $\Omega_{\chi^o} h^{2} > 1.1$. The remaining areas are excluded for the reasons noted.}
	\label{fig:OPM_mgvstanbSSC}
\end{figure}

Next, a full scan of the parameter space is performed for the dilaton scenario for $m_{1/2},~ m_{0} = \frac{1}{\sqrt{3}}m_{1/2},~A_{0} = -m_{1/2},$ and taking tan$\beta$ as a free parameter. With $m_{t} = 171.4$ GeV, a small region of the parameter space which satisfies
all constraints is found.  We exhibit the parameter space
which results in a relic density satisfying the WMAP limits in Fig.~\ref{fig:OPM_mgvstanb}. If the relic neutralino LSP comprises a sub-dominant component of the dark matter, we should not impose the lower bound on the WMAP limits. 
Thus, the parameter space for the four cases considered are 1) 0.1109 $\leq \Omega_{\chi^o} h^{2} \leq$ 0.1177, 2)
0.095 $\leq \Omega_{\chi^o} h^{2} \leq$ 0.129, 3) 0 $\leq \Omega_{\chi^o} h^{2} \leq$ 0.1177, and 4) 0 $\leq \Omega_{\chi^o} h^{2} \leq$ 0.129. Note that the parameter space shown in Fig.~\ref{fig:OPM_mgvstanb} is allowed by all constraints, except those regions noted. Fig.~\ref{fig:OPM_mgvstanbSSC} also displays the parameter space allowed by all constraints, though the parameter space for the SSC model of the dark matter density is shown in totality.

Imposing all the experimental constraints, we find that the viable WMAP parameter space is in the range tan$\beta$ = 35.2 to tan$\beta$ = 38 as shown in Fig.~\ref{fig:OPM_mgvstanb}. Extending the dark matter density to the SSC model as shown in Fig.~\ref{fig:OPM_mgvstanbSSC}, the allowed parameter space expands from tan$\beta$ = 10.2 to tan$\beta$ = 38. We show in Table~\ref{tab:AST} the allowed ranges for the CP-even Higgs boson mass satisfying all constraints for each range of $\Omega_{\chi^o} h^{2}$ within the parameter space. As we see, the range of the Higgs mass is highly constrained in each of the cases. For the superpartner spectra allowed by the constraints, there are only two hierarchal mass patterns of the four lightest sparticles:  1) $\widetilde{\chi}_{1}^{0} < \widetilde{\tau} < \widetilde{e}_{R} < \widetilde{\chi}_{1}^{\pm}$, and 2) $\widetilde{\chi}_{1}^{0} < \widetilde{\tau} < \widetilde{e}_{R} < \widetilde{\nu}_{\tau}$. The stau is NLSP in each case.  A characteristic of this coannihilation region is the nearly degenerate mass of the lightest neutralino and the stau, which is in fact what we find for these spectra in the WMAP region. The LSP in both the WMAP and SSC regions allowed by the constraints is found to be Bino-like.

\begin{table}[t]
	\centering
	\caption{Allowed ranges of the CP-even Higgs boson mass (in GeV) which satisfy the WMAP and SSC dark matter density limits as well as all other constraints.}
		\begin{tabular}{|c|c|} \hline
		$\Omega_{\chi^o} h^{2}$ & $m_{h}$~(GeV) \\ \hline\hline
		$0.1109$ $\leq \Omega_{\chi^o} h^{2} \leq$ $0.1177$  &  $117.17$~-~$118.58$\\ \hline
		$0.095$ $\leq \Omega_{\chi^o} h^{2} \leq$ $0.129$  &  $117.16$~-~$118.60$\\ \hline
		$0$ $\leq \Omega_{\chi^o} h^{2} \leq$ $0.1177$  &  $116.90$~-~$118.64$\\ \hline
		$0$ $\leq \Omega_{\chi^o} h^{2} \leq$ $0.129$  &  $116.90$~-~$118.64$\\	\hline
		$0$ $\leq \Omega_{\chi^o} h^{2} \leq$ $1.1$  &  $114.00$~-~$118.64$\\	\hline
		\end{tabular}
		\label{tab:AST}
\end{table}

It has been shown in~\cite{LHCtwotau, LHCthreetau} that it is possible to obtain mass measurements of the supersymmetric particles in the neutralino-stau coannihilation region  by utilizing each final state and parameterizing kinematical observables, such as those described in the previous section, in terms of the SUSY masses. The goal of such an analysis would be to determine the mSUGRA model parameters $m_0$, $m_{1/2}$, $A_0$, and tan$\beta$ since we want to determine the dark matter content and the neutralino-proton cross section, while the fifth mSUGRA model parameter, $\mathrm{sign}(\mu)$, is assumed to be positive, since this is preferred by measurements of the $b\rightarrow s\gamma$ branching ratio and the muon $g_{\mu}-2$~\cite{MUON}.  To determine the mSUGRA parameters, we need four kinematical observables which are linearly independent functions of those parameters.  The determination of the parameters is then accomplished by inverting four such functional relationships.
An analysis of this type was discussed in~\cite{DARK}. 

The OPM parameter space satisfying all constraints which has been found is a subset of the minimal supergravity (mSUGRA) parameter space, whose collider signals has been the subject of much study over the years.  In particular, the allowed parameter space of OPM satisfying the WMAP constraints falls into the coannihilation regions of mSUGRA.  However, since this is a very small subset of the mSUGRA parameter space, the allowed superpartner spectra are somewhat constrained, and thus the possible signals of the model which might be observed at LHC are constrained as well.  Once there is experimental data
from LHC available, one may perform the analysis discussed in the previous paragraph to determine the mSUGRA model parameters.   
These may then be compared to the above OPM parameter space allowed by constraints. If the experimentally determined mSUGRA parameters happen to coincide with the allowed OPM parameter space, then this may provide an important clue to the structure of the underlying theory at high energy scales. One may also focus on the superpartner spectra of OPM and the resulting experimental signatures which should be observed at LHC if we live in a one-parameter model universe.  We give a generic discussion of the experimental signatures of OPM in the next section.    

\section{Generic Phenomenological Features and Possible Signatures of OPM at LHC} 

In a one-parameter model universe, predominantly squarks and gluinos will be produced at LHC. To discuss the possible phenomenology of the one-parameter model, we select one spectrum from each of the two regions of the parameter space: WMAP and SSC. The spectra are identified as the WMAP Sparticle Spectrum (WMAP SS), and the SSC Sparticle Spectrum (SSC SS). We can then analyze the probable channels and resulting signatures at LHC, and construct the opposite sign (OS) ditau invariant mass. The anticipated states decaying from the squarks and gluinos involve hadronic jets and tau, so the OS ditau invariant mass may provide some clues leading to discovery at LHC. To examine these probable channels, the cross-sections and branching ratios can then be calculated with PYTHIA~\cite{PYTH} and cross-checked with ISAJET~\cite{ISAJ}, using ISASUGRA to calculate the sparticle masses. Many analyses have been completed on the entire mSUGRA parameter space, the stau-neutralino coannihilation region in particular. We do not repeat these analyses, but focus on this much more constrained region of the mSUGRA parameter space predicted by the one-parameter model. 

\subsection{WMAP Sparticle Spectrum}

The WMAP parameter space for OPM is quite constrained by Eqs.~\ref{eqn:dilaton}. This defines the one-parameter model as a very constrained subset of mSUGRA. For a detailed analysis of potential LHC signals of mSUGRA in the context of CMS, see~\cite{CMSR}. Here we will examine the probable states within only the one-parameter model region of the mSUGRA parameter space. The WMAP SS selected is shown in Table II. 

\begin{table}[h]
	\footnotesize
	\renewcommand{\arraystretch}{1.0}
	\begin{center}
	\caption{Low energy supersymmetric particles and their masses (in GeV) for $m_{1/2} = 606$,
$m_{0} = 349.9$, $A_{0} = -606$, tan$\beta$ = 36, $\mu > 0$, $\Omega_{\chi^{o}} h^{2}$ =  0.1147.}
	\begin{tabular}{|c|c|c|c|c|c|c|c|c|c|}\hline

$h^0$ & $H^0$ & $A^0$ & $H^{\pm}$ & ${\widetilde g}$ & $\widetilde\chi_1^{\pm}$
& $\widetilde\chi_2^{\pm}$ & $\widetilde\chi_1^{0}$ & $\widetilde\chi_2^{0}$  \\ \hline

117.6  & 753.9   & 753.9    & 758.4   & 1377  & 482.9    & 840.8   & 254.1
&  483.0      \\ \hline

$\widetilde\chi_3^{0}$ & $\widetilde\chi_4^{0}$ & ${\widetilde t}_1$ & ${\widetilde t}_2$ &
${\widetilde u}_R/{\widetilde c}_R$ & ${\widetilde u}_L/{\widetilde c}_L$ &
${\widetilde b}_1$ & ${\widetilde b}_2$ & \\ \hline

832.4  & 840.4  & 932.6  & 1169 & 1251 & 1294 & 1109 & 1174 &\\ \hline

${\widetilde d}_R/{\widetilde s}_R$ & ${\widetilde d}_L/{\widetilde s}_L$ &
${\widetilde \tau}_1$ & ${\widetilde \tau}_2$ & ${\widetilde \nu}_{\tau}$ &
${\widetilde e}_R/{\widetilde \mu}_R$ & ${\widetilde e}_L/{\widetilde \mu}_L$ &
${\widetilde \nu}_e/{\widetilde \nu}_{\mu}$ & $LSP$\\ \hline

1246 & 1297 & 263.2 & 509.3 & 485.2 & 416.5 & 532.8 & 527.0 & \textit{Bino}\\
	\hline
	\end{tabular}
	\end{center}
\end{table}

The mass of the gluino is greater than the mass of the squarks, hence the allowed processes with the largest differential cross-sections are $q+q\rightarrow\widetilde{q}+\widetilde{q}$ and $q+g\rightarrow\widetilde{q}+\widetilde{g}$, where $\widetilde{q} = (\widetilde{q}_{L}, \widetilde{q}_{R})$. The largest cross-section is $q+q~\rightarrow~\widetilde{q}_{R}+\widetilde{q}_{R}$, with $\widetilde{q}_{R}~\rightarrow~q\widetilde{\chi}^{0}_{1}$ for $\widetilde{q}_{R}=(\widetilde{u}_{R}, \widetilde{d}_{R}, \widetilde{c}_{R}, \widetilde{s}_{R})$. The resulting signature is a high number of 2 jets events plus missing energy. The next largest cross-section is $q+q~\rightarrow~\widetilde{q}_{L}+\widetilde{q}_{L}$, where the branching ratio for $\widetilde{q}_{L}~\rightarrow~q\widetilde{\chi}^{\pm}_{1}$ is 65\% and the branching ratio for $\widetilde{q}_{L}~\rightarrow~q\widetilde{\chi}^{0}_{2}$ is 33\% for $\widetilde{q}_{L}=(\widetilde{u}_{L}, \widetilde{d}_{L}, \widetilde{c}_{L}, \widetilde{s}_{L})$. Therefore, $\widetilde{q}_{L}$ will decay to either a $\widetilde{\chi}^{\pm}_{1}$ or $\widetilde{\chi}^{0}_{2}$. The lightest chargino decays to a stau by $\widetilde{\chi}_{1}^{\pm}~\rightarrow~\widetilde{\tau}_{1}^{\pm}\nu_{\tau}$. The second lightest neutralino decays to a stau through $\widetilde{\chi}_{2}^{0}~\rightarrow~\widetilde{\tau}_{1}^{\mp}\tau^{\pm}$. The probability of either a $\widetilde{\chi}_{1}^{\pm}$ or $\widetilde{\chi}_{2}^{0}$ decaying to a $\widetilde{\tau}_{1}$ is essentially the same, and this can be attributed to the nearly degenerate mass between the $\widetilde{\chi}_{1}^{\pm}$ and $\widetilde{\chi}_{2}^{0}$, as shown in Table II. The stau will always decay to tau and LSP via $\widetilde{\tau}_{1}^{\pm}~\rightarrow~\widetilde{\chi}_{1}^{0}\tau^{\pm}$. The process $q+q~\rightarrow~\widetilde{q}_{L}+\widetilde{q}_{R}$, which are just combinations of the above, has the next largest cross-section. To summarize the probable cascade decays for $q+q~\rightarrow~\widetilde{q}+\widetilde{q}$ where $\widetilde{q}=(\widetilde{u}, \widetilde{d}, \widetilde{c}, \widetilde{s})$, they are:

\begin{itemize}

  \item $\widetilde{q}_{R}~\rightarrow~\textit{q}\widetilde{\chi}_{1}^{0}$
  \item $\widetilde{q}_{L}~\rightarrow~\textit{q}\widetilde{\chi}_{1}^{\pm}~\rightarrow~q\widetilde{\tau}_{1}^{\pm}\nu_{\tau}~\rightarrow~q\nu_{\tau}\tau^{\pm}\widetilde{\chi}_{1}^{0}$
  \item $\widetilde{q}_{L}~\rightarrow~\textit{q}\widetilde{\chi}_{2}^{0}~\rightarrow~q\widetilde{\tau}_{1}^{\mp}\tau^{\pm}~\rightarrow~q\tau^{\pm}\tau^{\mp}\widetilde{\chi}_{1}^{0}$

\end{itemize}

As these processes show, combinations of these three channels will result in one tau, two tau, and three tau events with two hadronic jets, plus missing energy from the stable $\widetilde{\chi}_{1}^{0}$ LSP and tau neutrinos. These tau events will be well in excess of the observable limit as calculated in the next section, presenting the opportunity for clear distinction between the one-parameter model region of the mSUGRA parameter space and the background. Now we examine gluino decays. After the production of exclusively squarks, the next largest cross-sections are $q+g~\rightarrow~\widetilde{q}+\widetilde{g}$, where $\widetilde{q} = (\widetilde{q}_{L}, \widetilde{q}_{R})$. The heavier mass of the gluinos over the squarks in the one-parameter model requires the gluinos decay to squarks. The stop and sbottom are the lightest squarks, so these decays will be from gluinos to bottom and top squarks 73\% of the time. The remaining 27\% of the time the gluinos will decay to up, down, charm, and strange squarks. The branching ratios for the decay $\widetilde{g}~\rightarrow~\widetilde{b}_{1}\overline{b}$ is 20\% and $\widetilde{g}~\rightarrow~\widetilde{b}_{2}\overline{b}$ is 13\%, whereas $\widetilde{g}~\rightarrow~\widetilde{t}_{1}\overline{t}$ is 28\% and $\widetilde{g}~\rightarrow~\widetilde{t}_{2}\overline{t}$ is 12\%. Therefore, the $\widetilde{t}_{1}$ and $\widetilde{b}_{1}$ channels are most favored since $\widetilde{t}_{1}$ and $\widetilde{b}_{1}$ are lighter than $\widetilde{t}_{2}$ and $\widetilde{b}_{2}$. The top squark will decay via $\widetilde{t}_{1}~\rightarrow~t\widetilde{\chi}_{1}^{0}$ 41\% of the time, and $\widetilde{t}_{1}~\rightarrow~b\widetilde{\chi}_{1}^{\pm}$ 34\% of the time. The top quark decays to a b quark plus either jets or leptons through a $W^{\pm}$ boson. The bottom squark decays via $\widetilde{b}_{1}~\rightarrow~t\widetilde{\chi}_{1}^{-}$ 41\% of the time. To summarize the most probable results of the gluino cascade decays are

\begin{itemize}

\item $\widetilde{g}~\rightarrow~\widetilde{t}_{1}\overline{t}~\rightarrow~t\overline{t}\widetilde{\chi}_{1}^{0}$

\item $\widetilde{g}~\rightarrow~\widetilde{b}_{1}\overline{b}~\rightarrow~\overline{b}t\widetilde{\chi}_{1}^{-}~\rightarrow~\overline{b}t\widetilde{\tau}_{1}^{-}\nu_{\tau}~\rightarrow~\overline{b}t\tau^{-}\nu_{\tau}\widetilde{\chi}_{1}^{0}$

\item $\widetilde{g}~\rightarrow~\widetilde{q}_{R}\overline{q}~\rightarrow~q\overline{q}\widetilde{\chi}_{1}^{0}$

\end{itemize}

\noindent The combination of one of these gluino decays with one of the squark decays will produce one tau, two tau, and three tau events with two or more jets, plus missing energy. It is significant to notice that each $\widetilde{t}_{1}$ and $\widetilde{b}_{1}$ will produce a b-jet. Each stop and sbottom is accompanied by a top or bottom quark. Each top quark also produces a b-jet, so all b-jets will be produced in pairs, most in pairs of $b\overline{b}$. To emulate the expected LHC experience, PGS4 parameterizes b-tagging efficiency as a function of jet $P_{T}$. For our study here, we use a post-trigger level jet $P_{T}$ cut of 60 GeV. For jet $P_{T} >$ 60 GeV, the b-tagging efficiency in PGS4 varies from $\sim37\%$ to $\sim45\%$~\cite{LHC3}. Therefore, the number of events will decline for sequentially higher number of b-jets, and there will be more than three times as many one b-jet events as two b-jet events. For this reason, we will use the percentage of one b-jet events to understand the phenomenology of the one-parameter model, even though no single b-jets are produced. Examining the processes listed above, only the gluino decays result in a lepton as a result of the $W^{\pm}$ boson from the top quark, where $l = (e, \mu, \tau)$, therefore the number of tau events will encompass the majority of overall lepton events, in contrast to the low percentage of tau events per overall lepton events within the background. Namely, the large number of tau events in excess of the background are the most likely one-parameter model fingerprint. Hence, we conclude the most constructive analysis of the one-parameter model phenomenology is to study these specific collider signatures:

\begin{itemize}

\item 1 tau lepton, 1 tau and $\geq$1 b-jet, 2 tau leptons, 1 lepton, 2 leptons and $\geq$2 jets, 2 jets, 3 jets, 1 b-jet

\end{itemize}

\begin{figure}[t]
	\centering
		\includegraphics[width=0.45\textwidth]{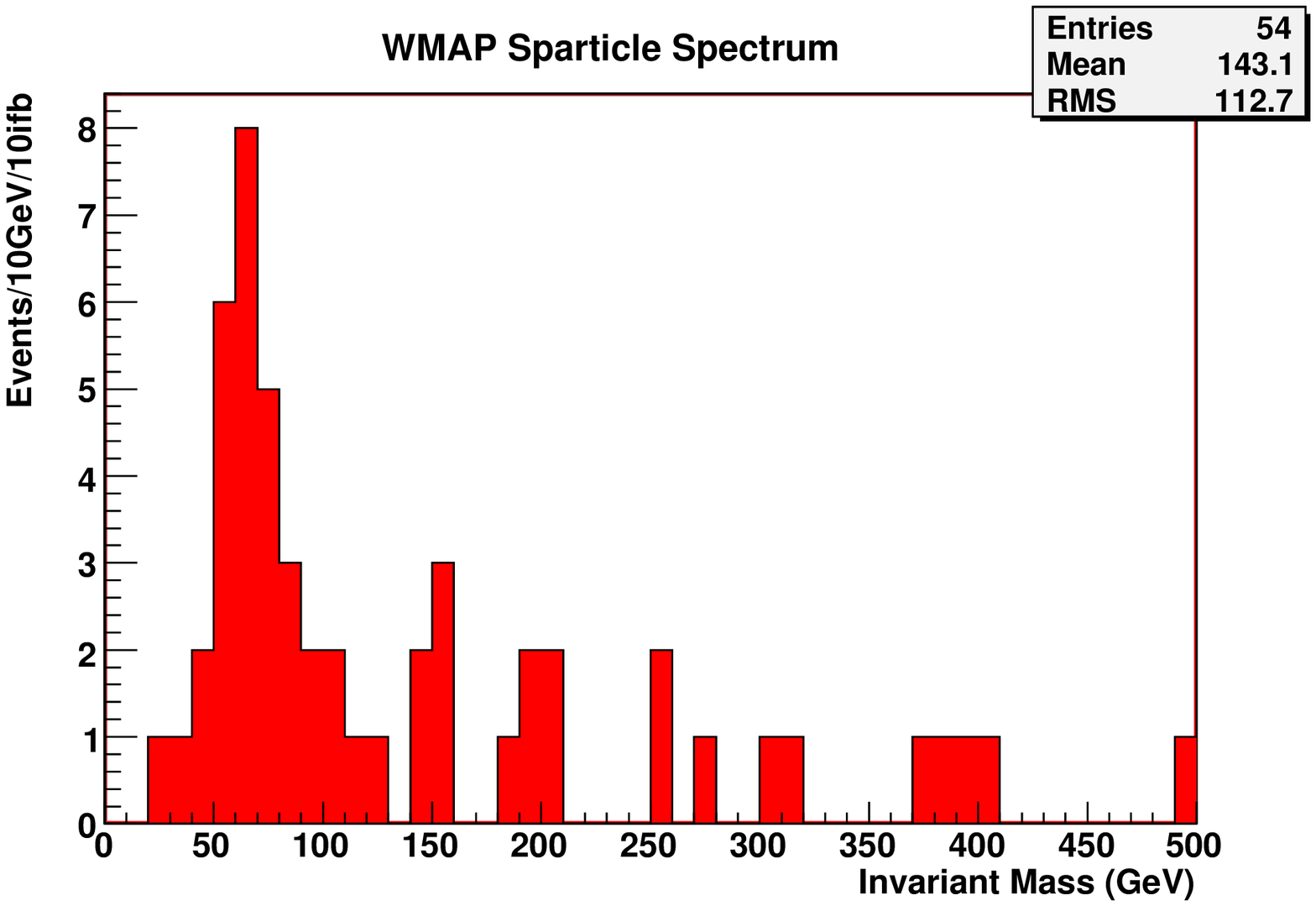}
		\includegraphics[width=0.45\textwidth]{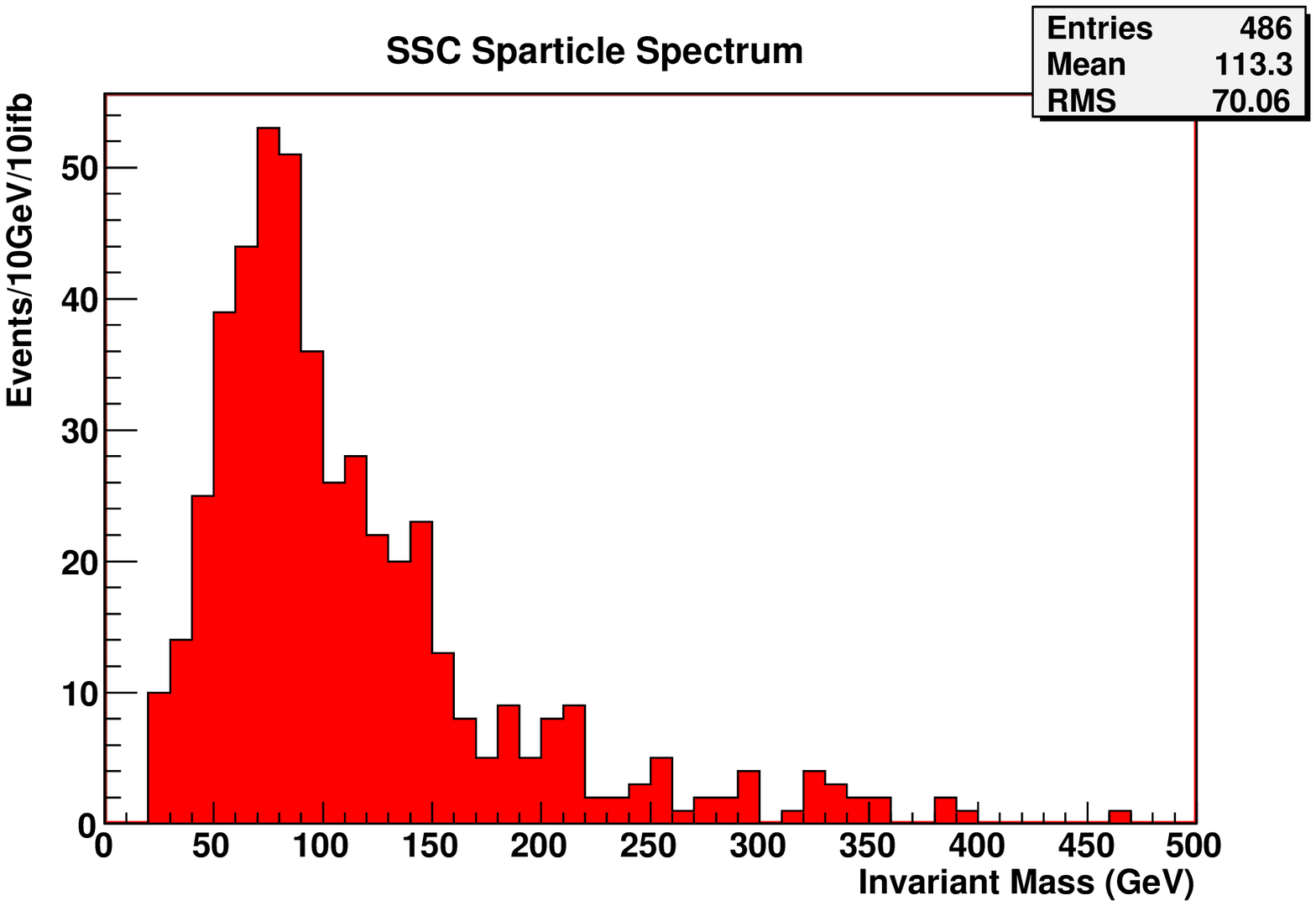}
	\caption{$\tau^{+}\tau^{-}$ invariant mass of the WMAP sparticle spectrum (in GeV), $m_{1/2} = 606$,
$m_{0} = 349.9$, $A_{0} = -606$, tan$\beta$ = 36, $\mu > 0$, $\Omega_{\chi^{o}} h^{2}$ =  0.1147, and the SSC sparticle spectrum (in GeV), $m_{1/2} = 475$, $m_{0} = 274.2$, $A_{0} = -475$, tan$\beta = 18$, $\mu > 0$, $\Omega_{\chi^{o}} h^{2}$ =  0.8496.}
	\label{fig:InvMass}
\end{figure}

\noindent In fact, we will use these signatures in the next section in our effort to compare the phenomenology of the one-parameter model with a different string vacua, that is, an intersecting $D$6-brane model. To conclude the analysis of the WMAP SS, we construct the $\tau^{+}\tau^{-}$ invariant mass for this WMAP spectrum in Fig.~\ref{fig:InvMass}. 

\subsection{SSC Sparticle Spectrum}

\begin{table}[h]
	\footnotesize
	\renewcommand{\arraystretch}{1.0}
	\begin{center}
	\caption{Low energy supersymmetric particles and their masses (in GeV) for $m_{1/2} = 475$,
$m_{0} = 274.2$, $A_{0} = -475$, tan$\beta = 18$, $\mu > 0$, $\Omega_{\chi^{o}} h^{2}$ =  0.8496.}
	\begin{tabular}{|c|c|c|c|c|c|c|c|c|c|}\hline

$h^0$ & $H^0$ & $A^0$ & $H^{\pm}$ & ${\widetilde g}$ & $\widetilde\chi_1^{\pm}$
& $\widetilde\chi_2^{\pm}$ & $\widetilde\chi_1^{0}$ & $\widetilde\chi_2^{0}$  \\ \hline

116.1   & 746.5   & 746.5    & 751.1   & 1100  & 372.6    & 689.3   & 196.0
&  372.6      \\ \hline

$\widetilde\chi_3^{0}$ & $\widetilde\chi_4^{0}$ & ${\widetilde t}_1$ & ${\widetilde t}_2$ &
${\widetilde u}_R/{\widetilde c}_R$ & ${\widetilde u}_L/{\widetilde c}_L$ &
${\widetilde b}_1$ & ${\widetilde b}_2$ & \\ \hline

679.1  & 689.0  & 733.4  & 973.4 & 999.7 & 1033 & 922.5 & 979.8 &\\ \hline

${\widetilde d}_R/{\widetilde s}_R$ & ${\widetilde d}_L/{\widetilde s}_L$ &
${\widetilde \tau}_1$ & ${\widetilde \tau}_2$ & ${\widetilde \nu}_{\tau}$ &
${\widetilde e}_R/{\widetilde \mu}_R$ & ${\widetilde e}_L/{\widetilde \mu}_L$ &
${\widetilde \nu}_e/{\widetilde \nu}_{\mu}$ & $LSP$\\ \hline

996.6 & 1036 & 294.3 & 418.4 & 403.4 & 327.7 & 419.6 & 412.2 & \textit{Bino}\\
	\hline
	\end{tabular}
	\end{center}
\end{table}

One representative spectrum was selected from the SSC region of the parameter space, and the masses are shown in Table III. The probable states do not vary from those of the WMAP region, though the branching ratios for the chargino and neutralino decays do vary. The processes with the largest cross-sections are the same as with the WMAP SS, that is, the production of squarks and then the production of squarks and gluinos. The only real difference involves the decay of charginos and neutralinos. The branching ratio for $\widetilde{\chi}_{1}^{\pm}~\rightarrow~\widetilde{\tau}_{1}^{\pm}\nu_{\tau}$ is 70\%, as opposed to 95\% for the WMAP SS. The masses of $\widetilde{\chi}_{1}^{\pm}$ and $\widetilde{\chi}_{2}^{0}$ are still nearly degenerate for the SSC SS spectrum, however, the masss difference between the $\widetilde{\chi}_{1}^{\pm}$ or $\widetilde{\chi}_{2}^{0}$ and the $\widetilde{\tau}_{1}$ is about 78 GeV as opposed to 220 GeV for the WMAP SS. This accounts for the smaller branching ratio for the SSC SS. For the same reason, the decay $\widetilde{\chi}_{2}^{0}~\rightarrow~\widetilde{\tau}_{1}^{\mp}\tau^{\pm}$ is now a little less likely with a branching ratio of 72\%, as opposed to 96\% for the WMAP SS. The other non-negligible decay process for the chargino is $\widetilde{\chi}_{1}^{\pm}~\rightarrow~W^{\pm}\widetilde{\chi}_{1}^{0}$, where the branching ratio for this is 29\% since the SSC SS has a lighter LSP than the WMAP SS. This process was negligible for the WMAP SS. The production of a higgs boson is now a little more probable at 24\% via $\widetilde{\chi}_{2}^{0}~\rightarrow~h_{0}\widetilde{\chi}_{1}^{0}$, whereas WMAP SS higgs production through $\chi_{2}^{0}$ was negligible. 
The branching ratios for the bottom and top squark decays are little changed, hence, the most probable processes remain the same as those for the WMAP SS. This is also true for the gluinos as well. Therefore, the signatures to study for the SSC are the same as those listed for WMAP. Fig.~\ref{fig:InvMass} plots the OS ditau invariant mass for the SSC SS, in addition to the WMAP SS invariant mass. The peak occurs about 10 GeV higher for the SSC SS, however, the main distinction is the number of events per 10 GeV per 10 $fb^{-1}$ of LHC data. The lighter sparticle spectrum of the SSC SS affords higher sparticle production than the WMAP SS for the same integrated luminosity.

\section{Signatures of OPM vs. Non-universality at LHC}

\begin{figure}[t]
	\centering
		\includegraphics[width=0.67\textwidth]{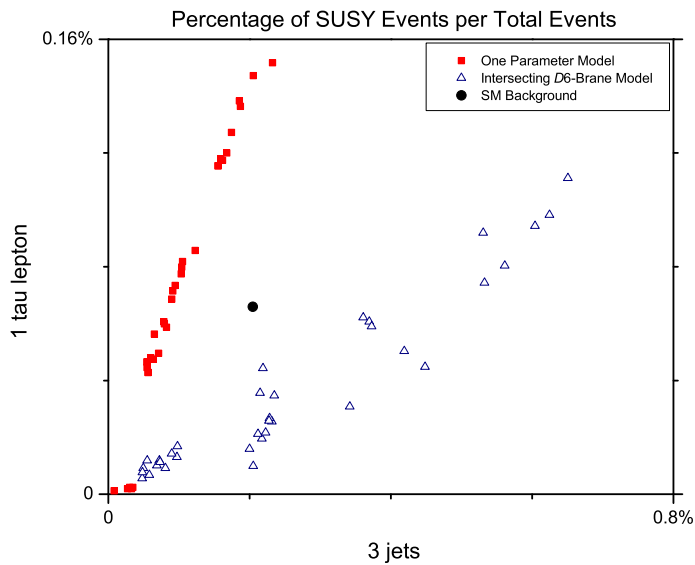}
	\caption{Percentage of 1 tau lepton vs. 3 jet events per 10 $\mbox{fb}^{-1}$ of integrated luminosity at LHC, for both
	the intersecting $D6$-brane model and the one-parameter model. The round marker indicates the observable limit due to the Standard Model background. Total events = signal + background.}
	\label{fig:OPMvsRM1}
\end{figure}

\begin{figure}[hf]
	\centering
		\includegraphics[width=0.67\textwidth]{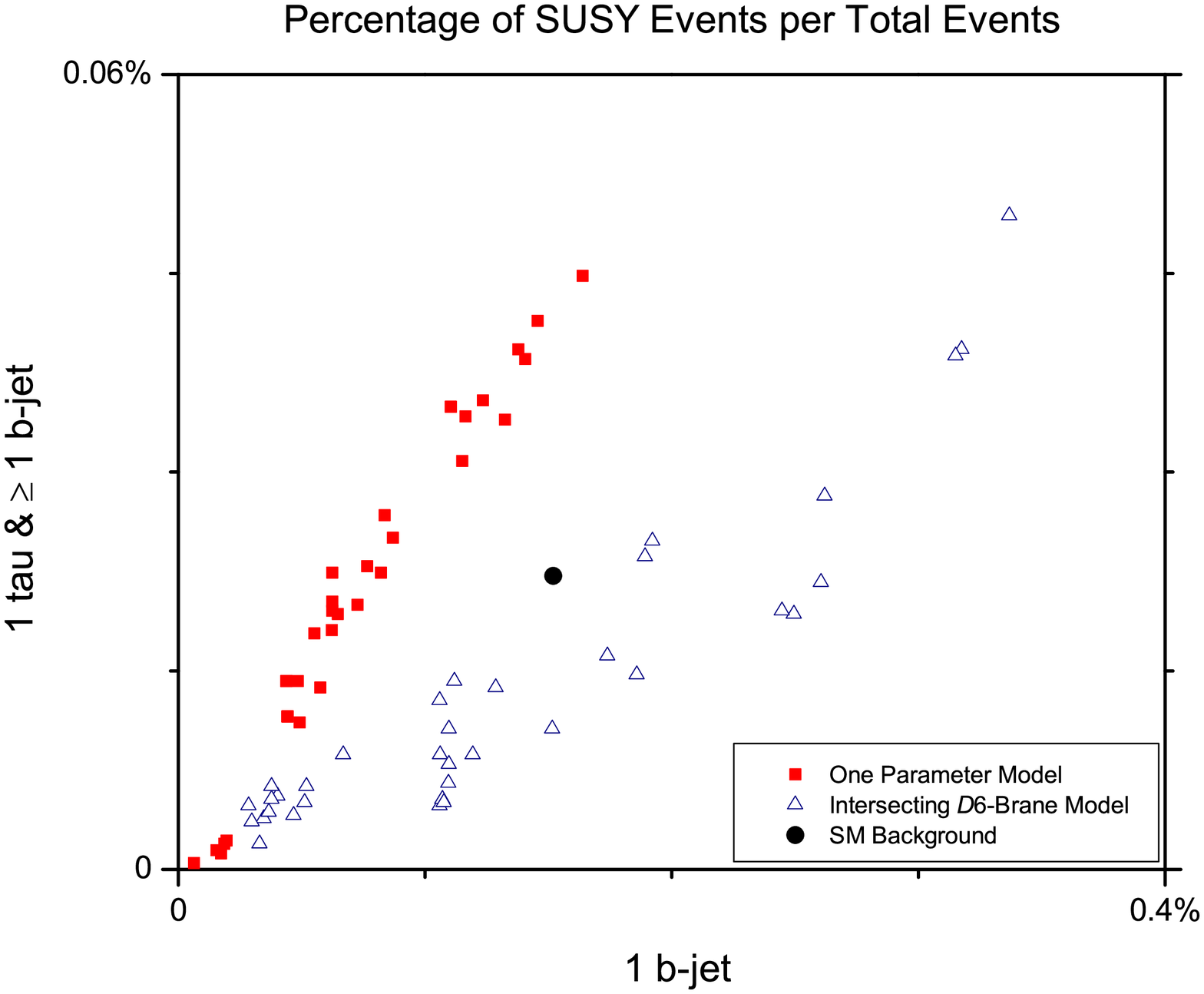}
	\caption{Percentage of 1 tau and $\ge$ 1 b-jet vs. 1 b-jet events per 10 $\mbox{fb}^{-1}$ of integrated luminosity at LHC, for both
	the intersecting $D6$-brane model and the one-parameter model. The round marker indicates the observable limit due to the Standard Model background. Total events = signal + background.}
	\label{fig:OPMvsRM2}
\end{figure}

\begin{figure}[hf]
	\centering
		\includegraphics[width=0.67\textwidth]{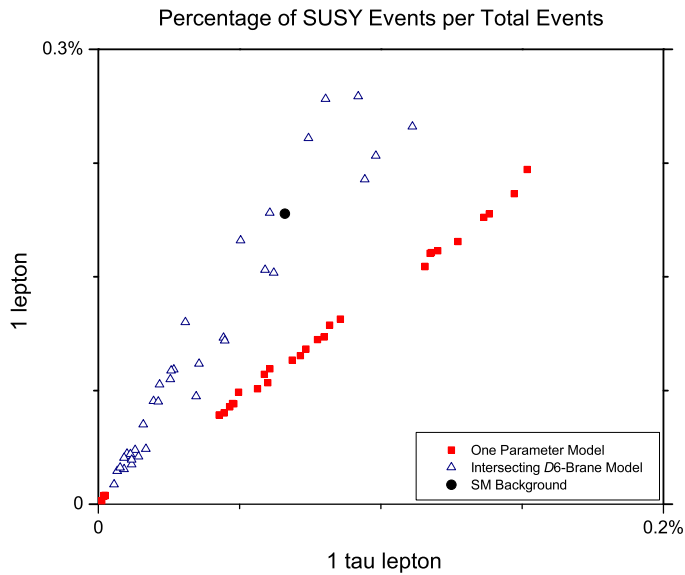}
	\caption{Percentage of 1 lepton vs. 1 tau lepton events per 10 $\mbox{fb}^{-1}$ of integrated luminosity at LHC, for both
	the intersecting $D6$-brane model and the one-parameter model. The round marker indicates the observable limit due to the Standard Model background. Total events = signal + background.}
	\label{fig:OPMvsRM3}
\end{figure}

In this section, we discuss the LHC signatures of the one-parameter model and compare them to those of an intersecting D6-brane model non-universal soft terms.  See~\cite{Feldman:2007fq} for a similar analysis.  To simulate events for different regions of the allowed parameter space, the superpartner mass spectra are first calculated using {\tt SuSpect 2.34}. Then production cross-sections and branching ratios are calculated using PYTHIA 6.4.14~\cite{PYTH}. The simulated events are then generated using the code PGS4~\cite{PGS4}. A SUSY Le Houches Accord (SLHA)~\cite{SLHA} file is output by Suspect 2.34 and this SLHA file is then called by PYTHIA via PGS4. In the PYTHIA card file, MSEL = 39 is used to generate 91 Minimal Symmetric Standard Model (MSSM) $2\rightarrow2$ production processes, excluding only single higgs production. The default configuration of the LHC Detector Card and the Level 1 (L1) trigger are used in PGS4. The L1 trigger level cuts are close to the actual values used by the Compact Muon Solenoid Detector (CMS) experiment. A table of the L1 trigger level cuts can be found here~\cite{L1TR}. A total integrated luminosity of 10 $fb^{-1}$ of data was simulated for all signatures. This corresponds to approximately the first few years of data collection at LHC. At this point, post-trigger level cuts over and above the L1 trigger level cuts can be applied to streamline the data even further. To apply post-trigger level cuts and count collider observables, the program Chameleon Root (ChRoot)~\cite{ROOT} was used. The post-trigger level cuts used for the one-parameter model are

\begin{itemize}
	\item $P_{T} >$ 60 GeV and $|\eta| <$ 3 for jets
	\item $P_{T} >$ 20 GeV and $|\eta| <$ 2.4 for photons and leptons
	\item $P_{T}^{miss} >$ 215 GeV for missing transverse momentum.
\end{itemize}

These same post-trigger level cuts are also applied to the Standard Model (SM) background. Constructing an estimate of the SM background is certainly nontrivial. The technical difficulty involves the number of background events. The number of background events can be six orders of magnitude larger than the signal, so most SM events must be discarded in the interest of compute time. Another major issue concerns simulating the largest component of the SM background, QCD physics. The simulation of W-bosons, quarks, and gluons is problematic. In the interest of reducing the compute time as much as possible, we use the SM background sample on the LHC Olympics website~\cite{LHCO}. This background sample was used for the LHC Olympics and contains 5 $fb^{-1}$ of LHC SM background data. We utilize this sample to formulate an estimate of the SM background for an integrated luminosity of 10 $fb^{-1}$ of detector data. This SM background sample contains dijets, \textit{t}$\overline{t}$, and W/Z+jets processes. To determine if a signal is observable above the SM background, an inclusive count of the individual signatures in the signal is compared to a count of the individual signatures in the background. In order for a signature to be observable above the background, the following statistical requirements must be satisfied~\cite{KANE}:
\begin{eqnarray}
 \centering
	&\frac{S}{\sqrt{B}} > 4, \ \ \ \ \ \ \ \ \ \ \ \ \  S > 5&\nonumber
\end{eqnarray}
where S is the number of signal events and B is the number of background events that survive the trigger level and post-trigger level cuts.
An estimate of the observable limit due to the SM background can be computed and compared to the MSSM production process number of events for the signature in order to determine whether a particular signature is observable above the background after all cuts have been applied. In our analysis, the signal is composed of SUSY signatures involving leptons (e,~$\mu$,~and $\tau$), jets, and b-tagged jets. 

Much of the older work toward constructing semi-realistic string vacua was 
done in the context of heterotic strings.  In particular, many of the most
phenomenologically interesting models were those constructed within the free-fermionic
formulation~\cite{AEHN}, and it is really from these types of models that the one-parameter model
was first defined~\cite{Lopez:1993rm, Lopez:1994fz, Lopez:1995hg}.  As we have mentioned, the same basic structure
of the one-parameter model also arises in the context of heterotic M-theory constructions
as well as Type IIB orientifold flux compactifications.  Besides these types of 
string vacua, there are also recent constructions involving Type IIA compactifications involving $D6$-branes intersecting
at angles. Such models have been the subject of much study in recent years, and we refer the reader to
\cite{Blumenhagen:2005mu, Blumenhagen:2006ci} for recent reviews.   The soft terms
for intersecting $D6$-branes are in general non-universal~\cite{Kors:2003wf}, in contrast in the 
one-parameter model as well as the standard framework, mSUGRA.  Thus, it is an
interesting question whether or not it is phenomenologically possible to distinguish between these two
different types of string compactifications by what is observed at LHC.

\begin{figure}[hf]
	\centering
		\includegraphics[width=0.67\textwidth]{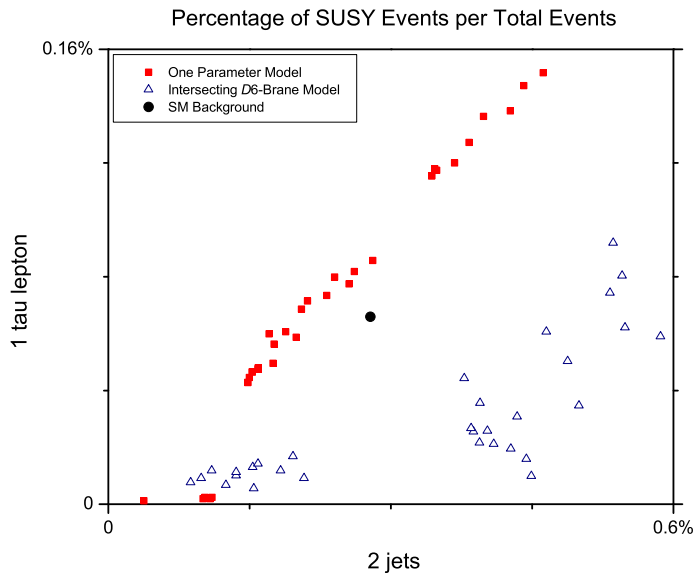}
	\caption{Percentage of 1 tau lepton vs. 2 jet events per 10 $\mbox{fb}^{-1}$ of integrated luminosity at LHC, for both
	the intersecting $D6$-brane model and the one-parameter model. The round marker indicates the observable limit due to the Standard Model background. Total events = signal + background.}
	\label{fig:OPMvsRM4}
\end{figure}

\begin{figure}[hf]
	\centering
		\includegraphics[width=0.67\textwidth]{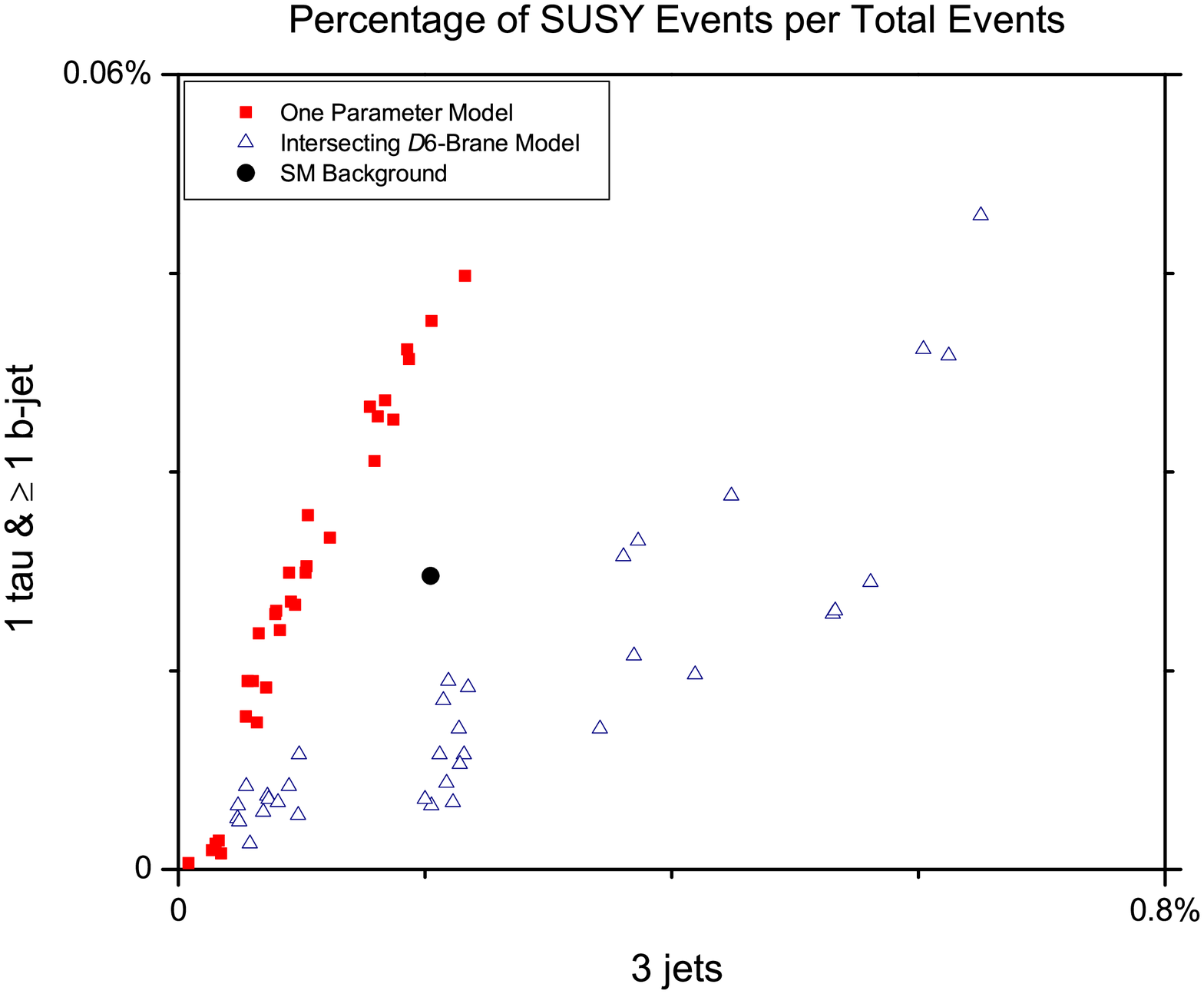}
	\caption{Percentage of 1 tau and $\ge$ 1 b-jet vs. 3 jet events per 10 $\mbox{fb}^{-1}$ of integrated luminosity at LHC, for both
	the intersecting $D6$-brane model and the one-parameter model. The round marker indicates the observable limit due to the Standard Model background. Total events = signal + background.}
	\label{fig:OPMvsRM5}
\end{figure}

\begin{figure}[hf]
	\centering
		\includegraphics[width=0.67\textwidth]{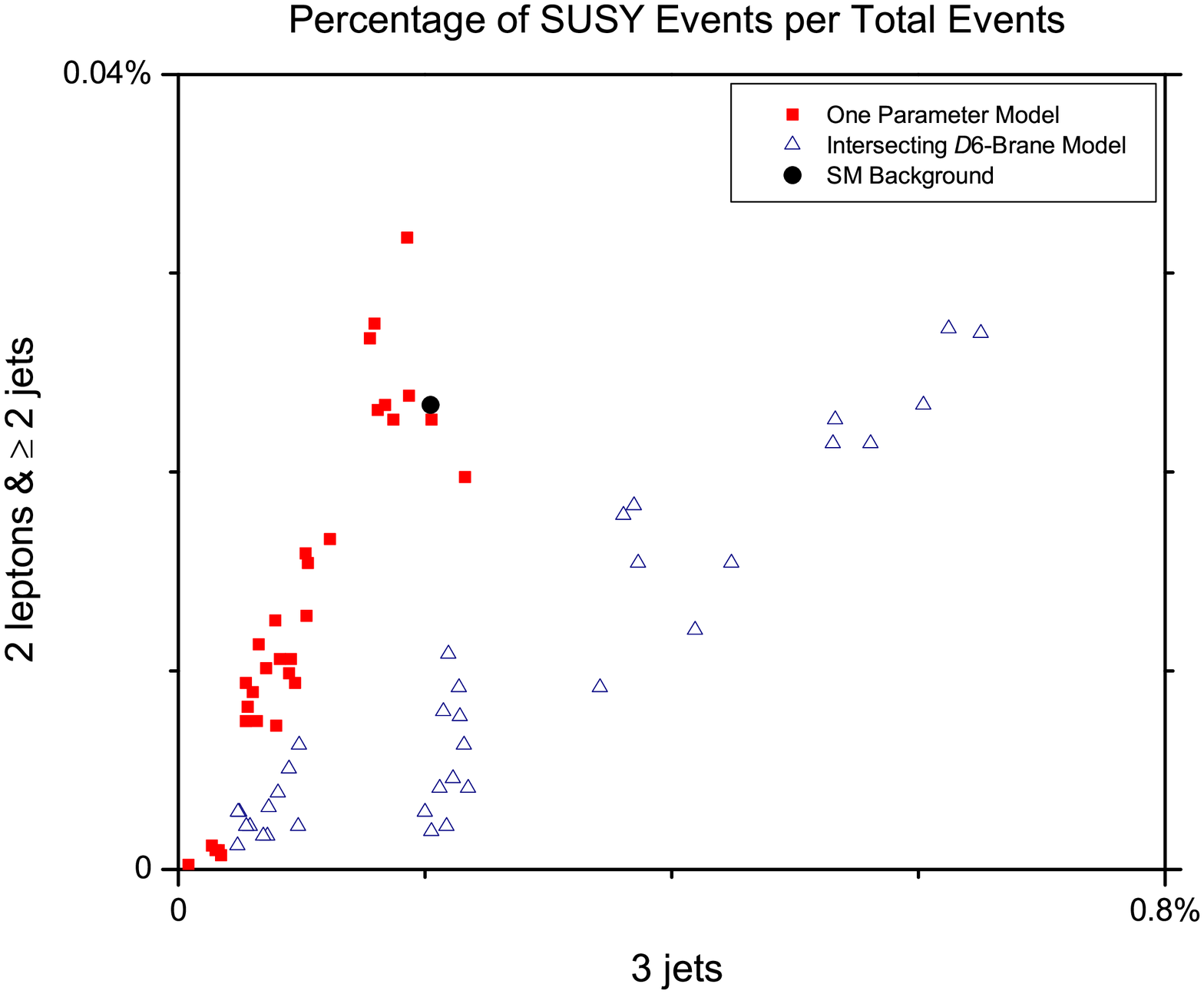}
	\caption{Percentage of 2 leptons and $\ge$ 2 jets vs. 3 jet events per 10 $\mbox{fb}^{-1}$ of integrated luminosity at LHC, for both
	the intersecting $D6$-brane model and the one-parameter model. The round marker indicates the observable limit due to the Standard Model background. Total events = signal + background.}
	\label{fig:OPMvsRM7}
\end{figure}

\begin{figure}[hf]
	\centering
		\includegraphics[width=0.67\textwidth]{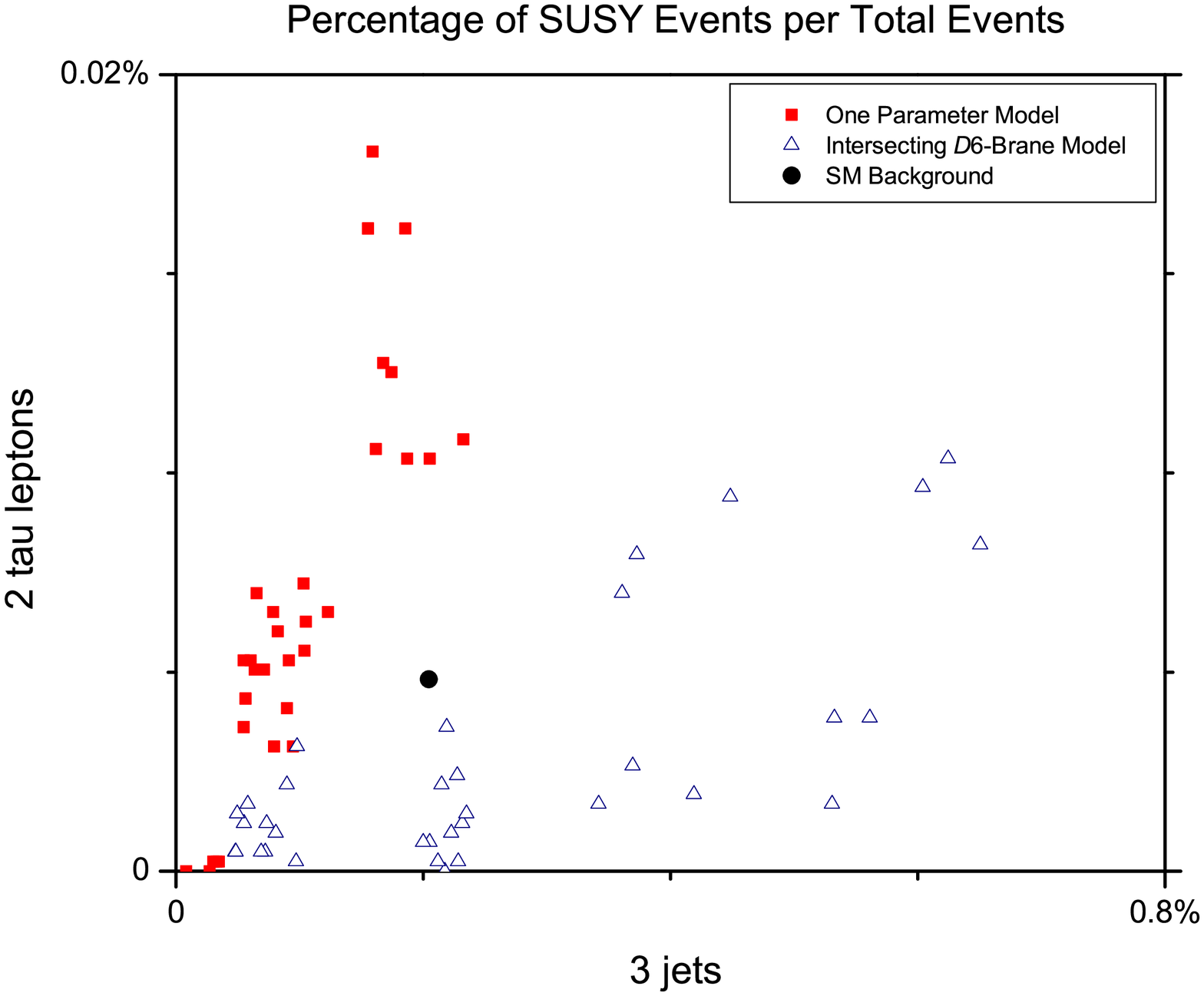}
	\caption{Percentage of 2 tau leptons vs. 3 jet events per 10 $\mbox{fb}^{-1}$ of integrated luminosity at LHC, for both
	the intersecting $D6$-brane model and the one-parameter model. The round marker indicates the observable limit due to the Standard Model background. Total events = signal + background.}
	\label{fig:OPMvsRM8}
\end{figure}

In ref.~\cite{RIBM} an explicit example of a supersymmetric intersecting $D6$-brane model in Type IIA string
theory was constructed which possesses many desirable phenomenological properties.  In particular,
the model has three generations of SM fermions and also exhibits automatic gauge coupling unification at tree-level.  In
addition, it is possible to obtain correct masses and mixings for both up and down-type quarks as well as the 
tau lepton.  The soft supersymmetry breaking terms were also calculated for this model, and it was shown
that there are regions within the parameter space which may generate the observed dark matter density and 
superpartner spectra satisfying all presently known constraints.  Given the desirable phenomenology of this 
model, it provides a suitable candidate with which to compare the one-parameter model.  In particular, 
the question that we would like to address is if there are distinguishing characteristics in the collider signatures between this class of 
string vacua constructed with intersecting $D6$-branes in Type IIA and other string constructions mentioned earlier having
soft terms similar to OPM.   

As discussed in the previous section, there are signatures at LHC that could provide distinguishing characteristics of the one-parameter model. For the present, we want to investigate if these signatures differ between the one-parameter model and the intersecting $D$6-brane model. In Figs.4-10 we plot two of these collider signatures against each
other for both the one-parameter model and the intersecting $D6$-brane model. For each of the cases shown, there seems to be a clear separation between the one-parameter model and the intersecting $D6$-brane model. All spectra simulated in these figures from both models are within the WMAP region of the allowed parameter space. Namely, thirty-one spectra from the one-parameter model and thirty-five spectra from the intersecting $D6$-brane model were simulated in the event generator. Table~\ref{tab:masspatt} lists the different  patterns of mass hierarchies for the one-parameter model parameter space and the intersecting $D6$-brane model. The mass patterns of the thirty-five superpartner spectra for the intersecting $D6$-brane model simulated in the event generator are the {\it ID6BraneP1} chargino pattern. This is in contrast to the two different stau patterns of the one-parameter model, {\it OPMP1} and {\it OPMP2}. The thirty-one one-parameter model spectra simulated in the event generator were a combination of both the {\it OPMP1} and {\it OPMP2} stau mass patterns. The patterns in these charts corroborate the details given in the previous section that the number of tau events as a percentage of overall lepton events for the one-parameter model will be high. In fact, the percentage of tau events is much higher in the one-parameter model than in the intersecting $D$6-brane model. We will perform a complete phenomenological analysis of the intersecting $D$6-brane model in the future, but suffice it to say that by examining the mass pattern for {\it ID6BraneP1} in Table~\ref{tab:masspatt}, it is evident that the decay of the $\widetilde{\chi}_{1}^{\pm}$ or $\widetilde{\chi}_{2}^{0}$ to a $\widetilde{\tau}_{1}$ is no longer present due to the $\widetilde{\tau}_{1}$ now being heavier than the $\widetilde{\chi}_{1}^{\pm}$ and $\widetilde{\chi}_{2}^{0}$. The elimination of this channel alone will reduce the production of tau, as the charts illustrate. We defer a more in-depth study of the additional intersecting $D6$-brane mass patterns in Table~\ref{tab:masspatt} versus the one-parameter model for future work.

\begin{table}[t]
	\centering
	\caption{Mass patterns of spectra allowed by all constraints for the one-parameter model (OPM) and the intersecting $D6$-brane model (IBM).}
		\begin{tabular}{|c|c|c|c|} \hline
		$\textnormal{Model}$ & $\textnormal{Pattern No.}$ &  $\textnormal{Pattern Type}$ & $\textnormal{Mass Pattern}$ \\ \hline\hline
		$\textnormal{OPM}$  &  $\textnormal{\it{OPMP1}}$  &  $\textnormal{Stau}$  &  $~\widetilde{\chi}_{1}^{0} ~<~ \widetilde{\tau} ~<~ \widetilde{e}_{R} ~<~ \widetilde{\chi}_{1}^{\pm}~$\\ \hline
		$\textnormal{OPM}$  &  $\textnormal{\it{OPMP2}}$  &  $\textnormal{Stau}$  &  $~\widetilde{\chi}_{1}^{0} ~<~ \widetilde{\tau} ~<~ \widetilde{e}_{R} ~<~ \widetilde{\nu}_{\tau}~$\\ \hline\hline
		$\textnormal{IBM}$  &  $\textnormal{\it{ID6BraneP1}}$  &  $\textnormal{Chargino}$  &  $~\widetilde{\chi}_{1}^{0} ~<~ \widetilde{\chi}_{1}^{\pm} ~<~ \widetilde{\chi}_{2}^{0} ~<~ \widetilde{\tau}~$\\ \hline
		$\textnormal{IBM}$  &  $\textnormal{\it{ID6BraneP2}}$  &  $\textnormal{Chargino}$  &  $~\widetilde{\chi}_{1}^{0} ~<~ \widetilde{\chi}_{1}^{\pm} ~<~ \widetilde{\tau} ~<~ \widetilde{\chi}_{2}^{0}~$\\	\hline
		$\textnormal{IBM}$  &  $\textnormal{\it{ID6BraneP3}}$  &  $\textnormal{Chargino}$  &  $~\widetilde{\chi}_{1}^{0} ~<~ \widetilde{\chi}_{1}^{\pm} ~<~ \widetilde{\tau} ~<~ \widetilde{e}_{R}~$\\	\hline
		$\textnormal{IBM}$  &  $\textnormal{\it{ID6BraneP4}}$  &  $\textnormal{Stau}$  &  $~\widetilde{\chi}_{1}^{0} ~<~ \widetilde{\tau} ~<~ \widetilde{\chi}_{1}^{\pm} ~<~ \widetilde{\chi}_{2}^{0}~$\\	\hline
		$\textnormal{IBM}$  &  $\textnormal{\it{ID6BraneP5}}$  &  $\textnormal{Stau}$  &  $~\widetilde{\chi}_{1}^{0} ~<~ \widetilde{\tau} ~<~ \widetilde{e}_{R} ~<~ \widetilde{\chi}_{1}^{\pm}~$\\ \hline
		\end{tabular}
		\label{tab:masspatt}
\end{table}

\section{Conclusion}

We have updated and surveyed the allowed parameter space of the one-parameter model. Our motivation for studying this model stems from the commonality of the universal soft supersymmetry breaking ansatz across multiple types of string compactifications. These include weak coupled and heterotic M-theory vacua, as well as Type IIB flux vacua, in particular the so-called large-volume compactification models.  By performing a comprehensive scan of the entire parameter space and filtering the results according to experimental constraints, the allowed parameter space was obtained. In the strict moduli dominant case, we found that there is no parameter space which can satisfy all of the constraints, whereas there is a small parameter space allowed for the dilaton scenario.  We identified the probable squark and gluino interactions and presented cascade decay modes that will produce specific favorable collider signatures. The dominant component of these favorable signatures are tau and hadronic jets. In future work, we plan to study the observable signatures at LHC for this model in somewhat more depth.  

We compared the collider signatures of the one-parameter model to a model with non-universal soft terms, in particular an intersecting $D6$-brane model with interesting phenomenological properties. We
found that for one particular intersecting $D$6-brane pattern of mass hierarchies, there are possible distinguishing characteristics between these two classes of models. Although there may also be a lot of overlap in the observable signatures of these two models, there are regions of the parameter space of each class which may give strikingly different observable signatures which may be used to distinguish them. Thus, it may be possible for LHC to say something about the structure of the underlying theory at high-energies, e.g. universality vs. non-universality.   We plan to investigate this possibility more deeply in an upcoming paper.

\section{Acknowledgments}

J.M. would like to thank David Toback, Bhaskar Dutta, and Teruki Kamon for useful discussions. This research was supported in part by the Mitchell-Heep Chair in High Energy Physics (CMC), by the Cambridge-Mitchell Collaboration in Theoretical Cosmology, and by the DOE grant DE-FG03-95-Er-40917.

\end{document}